# Designing as Construction of Representations:
# A Dynamic Viewpoint in Cognitive Design Research

Willemien Visser[1]
*INRIA - National Institute for Research in Computer Science and Control*

## ABSTRACT

This article presents a cognitively oriented viewpoint on design. It focuses on cognitive, dynamic aspects of real design, i.e., the actual cognitive activity implemented by designers during their work on professional design projects. Rather than conceiving designing as problem solving—Simon's symbolic information processing (SIP) ap-proach—or as a reflective practice or some other form of situated activity—the situativity (SIT) approach—we consider that, from a cognitive viewpoint, designing is most appropriately characterised as a construction of representations. After a critical discussion of the SIP and SIT approaches to design, we present our viewpoint. This presentation concerns the evolving nature of representations regarding levels of abstraction and degrees of precision, the function of external representations, and specific qualities of representation in collective design. Designing is described at three levels: the organisation of the activity, its strategies, and its design-representation construction activities (different ways to generate, transform, and evaluate representations). Even if we adopt a "generic design" stance, we claim that design can take different forms depending on the nature of the artefact, and we propose some candidates for dimensions that allow a distinction to be made between these forms of design. We discuss the potential specificity of HCI design, and the lack of cognitive design research occupied with the quality of design. We close our discussion of representational structures and activities by an outline of some directions regarding their functional linkages.

---

[1] Willemien Visser is a cognitive psychologist with an interest in cognitive design research. She is a senior researcher in the "EIFFEL—Cognition & Cooperation in Design" team of INRIA, the French National Institute for Research in Computer Science and Control.

## CONTENTS

# 1. INTRODUCTION

Design is an important, all-pervading domain of human activity, and the object of design is not limited to new sport cars but also covers artefacts as diverse as meals (Byrne, 1977), traffic signals (Bisseret, Figeac-Letang, & Falzon, 1988), route plans (Chalmé, Visser, & Denis, 2004; Hayes-Roth & Hayes-Roth, 1979), and, of course, software (Détienne, 2002).[12] This article compares the two main paradigms that govern today's cognitively oriented studies of design—abbreviated into "cognitive design studies"—and then presents our view on design, which considers that design is most appropriately characterised as a construction of representations.

## 1.1. Overview of the Article

The article is organised as follows. Through some preliminary terminological points, Section 1 presents the aspects of design on which we focus. Section 2 describes the two main paradigms in current cognitive design research (i.e., the symbolic information processing [SIP] and situativity [SIT] perspectives). Section 3, which comprises the main body of this article, describes the approach to design that we have been developing for some 20 years. Much of the data comes from our own work, which has largely consisted of empirical observational field studies conducted on professional designers in various task domains (software, mechanical, and industrial design). In the Conclusion, we review the central themes of this article and discuss some issues to be examined in further research.

## 1.2. Some Preliminary Terminological Points

This article focuses on cognitive, dynamic aspects of real design (i.e., the actual cognitive activity implemented by designers during their work on professional design projects). It is hoped that the elements presented will be used to build a descriptive process model.

This section makes explicit some presuppositions underlying these qualifications that are familiar to researchers from cognitive psychology and ergonomics, but which may be informative for readers from other disciplines.

"Cognitive aspects" refer to designers' cognitive processes, knowledge, and representations as distinguished from sociocultural and emotional aspects of design.

"Dynamic" refers to the use of knowledge and representations, in contrast to their content.

"Cognitive activity" refers to the way in which people realise their task at a cognitive level. "Task" refers to, either what people are supposed to do (prescribed task), or the task that they set themselves and which they carry out (actual task; Leplat, 1981).

We focus on *real* design rather than design performed in artificially restricted conditions (such as in laboratory experiments), and on *actual* design rather than the design task or the design process that are the reference in design methods, and prescriptive and stage models. In the rest of this article, "design activity" or "designing" refers to this real, actual design activity.

Design and Problem Solving

We consider that design *involves* problem solving, but that design *is* not (only) problem solving. This position will become clearer in what follows. Here we briefly discuss the notions "problem" and "problem solving," which seem to require some comment, given certain remarks and discussions in the design literature.

Problem

Much misunderstanding and disagreement seem to be due to the specific, technical sense of *problem* in cognitive psychology. Both in prescientific and in scientific contexts, the term is often used as referring to a *difficulty* or a *deficiency* (Kelley & Hartfield, 1996; Stolterman, 1991), whereas in cognitive psychology, "problem" qualifies someone's mental representation of their task (also noticed by Stacey & Eckert, 2003, p. 179).

We adopt Mayer's (1989) distinction between two types of problems. A task is a "routine problem" for someone if its representation, although not eliciting a memorised answer, can be solved by applying a well-known procedure. For "nonroutine problems," a person's task representation does not evoke a procedure, and so one must be constructed. For most adults, 863 • 725 will be a routine problem, whereas it will be a nonroutine problem for nearly every young child. The problem character of a task is thus a relative feature that depends on the task situation and on the person who must deal with it: A task that constitutes a problem for one person does not necessarily do so for another person (Leplat, 1981).

Design tasks involve solving both routine and nonroutine problems. This article focuses on the nonroutine problems in design projects. From the cognitive viewpoint that we adopt, these are the most interesting.

Problem Solving

Problem solving refers to a global process, not to one particular step. This process involves all the activities that lead from a problem specification to its solution(s). It takes place through a large number of small steps, "from problem detection through various attempts to problem solution or problem abandonment" (Gilhooly, 1989, p. 5). Examples of these steps are problem forming, problem finding (Simon, 1987/1995; see also Thomas, 1989), problem formulation (Gilhooly, 1989; Thomas, 1989), problem setting (Schön, 1983), problem structuring, and problem framing (Schön, 1988).

## 2. TWO ALTERNATIVE APPROACHES TO DESIGN

The two major approaches in cognitive design research are the classical cognitive-science paradigm in terms of symbolic information processing (SIP) and the main alternative approach in terms of situatedness and reflective practice.

---

[2] In this article, an "artefact" (something that is "man-made as opposed to natural", Simon, 1969/1999) refers to any entity that one may design be it physical (buildings or machine-tools) or symbolic (music, software, social welfare policies, route plans, or any procedure).

## 2.1. The Classical View of Design: The SIP Approach

The SIP approach to design is mainly based on Herbert A. Simon's work. With more or less profound modifications, its framework has been adopted by many authors (Akin, 1986; Baykan, 1996; Eastman, 1969; Goel, 1995). Si-mon's ideas continue to be "a dominant force within the field," as noticed by Roozenburg and Dorst (1999). They illustrate their claim by observing that in the first two Design Thinking Research Symposia in 1992 and 1994 (Cross, Christiaans, & Dorst, 1996; Cross, Dorst, & Roozenburg, 1992), "Simon was referred to more than anyone else: 31 direct references and goodness knows how many indirect ones in 32 papers" (Roozenburg & Dorst, 1999, p. 34).

Simon's bibliography, which comprises nearly 1000 titles,[3] contains only some 10 references directly discussing cognitive aspects of design. Yet, one of Simon's seminal works, *The Sciences of the Artificial* (1969/1999),[4] and his article *The Structure of Ill-Structured Problems* (1973/1984), are among the central references in cognitive design research.

Two steps can be distinguished in Simon's contribution to a cognitive design theory. First, together with Alan Newell, Simon elaborated the principles underlying the approach to problem solving that has since been called *SIP* (Newell & Simon, 1972; for a succinct presentation, Simon, 1979). Afterward, Simon applied this paradigm to design (Simon, 1969/1999, 1971/1975, 1973/1984, 1987/1995). In these analyses, Simon identified and elaborated several of the characteristics of design that have formed the basis of cognitive design research by later researchers.

### Simon's Framework for Design: *The Sciences of the Artificial*

In this section we first present four characteristics introduced by Simon that still occupy a central place in cognitive design research—even if they do not remain uncontested.

*Design Problems: Ill-Structured Problems?* Even if Simon was one of the first authors to *discuss* design problems' structuredness, he did not consider it a specific characteristic of design. He considers that many problems often treated as well-structured problems (WSP) would be better regarded as ill-structured problems (ISP; Simon, 1973/1984, p. 182). This even holds for chess playing and theorem proving, the kind of problems on which Newell and Simon (1972) based the analyses underlying their SIP model. These observations, however, do not lead Simon to conclude that all problems are ill-structured, or that problem structuredness is a relative characteristic. Instead, he considers that "ill-structured problems" are only so at first analysis: thanks to problem-solvers applying certain problem-solving strategies, these problems rapidly acquire structure.

---

[3] http://www.psy.cmu.edu/psy/faculty/hsimon/HSBib-1930–1950.htmlretrieved on September 15, 2005.
[4] This book went into three editions, each one completing and revising the previous one: 1969, 1981, and 1996. Our article refers to the third printing of the third edition (Simon, 1969/1999).

*Design: A Cognitive Activity Rather Than a Professional Status.* For ergonomists and cognitive psychologists, Simon is an intellectual precursor when he states, in 1969, that "design" refers to a type of cognitive activity, not a professional status restricted to certain design professionals, such as engineering designers or architects. "Everyone designs who devises courses of action aimed at changing existing situations into preferred ones" (Simon, 1969/1999, p. 111).

*Design: A Satisficing Activity.* Making decisions without complete information by accepting "good enough" solutions rather than optimising and calculating *the* optimum *one* (Simon, 1971/1975), is a distinctive characteristic of design.

*Design: A Regular Problem-Solving Activity.* For Simon, design *is* problem solving, exactly like the other activities he studied. According to his "nothing special" position (Klahr & Simon, 2001, p. 76), the "mechanisms that have shown themselves to be efficacious for handling large, albeit *apparently* well-structured domains should be extendable to ill-structured domains without any need for introducing qualitatively new components" (Simon, 1973/1984, p. 197). Solving ISPs "merely" requires a preliminary stage: one first structures the ISP, and then solves the resulting WSP. Simon's only remark about the relative importance of this structuring activity is that "there is merit to the claim that much problem solving effort is directed at structuring problems, and only a fraction of it at solving problems once they are structured" (p. 187).

Discussion of the SIP Approach

In this section, we discuss six aspects of design for which we consider that Simon misrepresented design (for a detailed discussion of these points, see Visser, 2004). We will end this section with some words on more nuanced positions formulated by Simon and a hypothesis concerning his view on design.

*Underestimating the Specificity of Ill-Structured Problems Solving.* As noticed previously, Simon (1973/1984) considers that, one first structures a problem and then solves it, using generally applicable problem-solving mechanisms. Empirical observations on professional designers at work have shown, however, that it is throughout the design process that these so-called "structuring" activities occur, by, for example, (re)interpretation, inference, exploration, and analogical reasoning.

*Underestimating the Importance of Problem-Representation Construction.* While it is true that Simon sometimes states that the construction of representations is important in design, he generally does so in a final subsection or presents it as a detail that "deserves further examination." It is clearly peripheral to an SIP approach to problem solving.

*Underestimating the Specificity of Nondeterministic Leaps—Overesti-mating the Importance of Recognition.* Simon does not deny that intuition, insight, and inspiration may play a role in expert activities. However, these activities are not the result of any mysterious, inexplicable internal force. "Most intuitive leaps are acts of recognition" (Simon, 1969/1999, p. 89).

"Interesting" design ideas often depend on leaps *between* domains (Visser, 1991). In this case, the number of possible candidates for recognition is huge and the chance that there already exist links between target and source representations is tiny. The required "nondeterministic" leaps may depend on profound analogies (Johnson-Laird, 1989, p. 313). The critical step then consists in designers "seeing" their target in such a way that their representation evokes an analogous source in memory! Once such a representation has been constructed, it is indeed "simply" an act of recognition.

*Overestimating the Importance of Systematic Problem Decomposition.* For Simon, systematic problem decomposition plays a fundamental role in design. Empirical studies show, however, that designers certainly decompose, but often not in a systematic way. The same design can often be broken down in different ways (Reitman, 1964, p. 296). Decomposition is frequently based, not on problem structure, but on a designer's experience (Goel & Pirolli, 1992). It often results in "unwanted side effects" (i.e., multiple interdependencies among subproblems that, as Simon (1973/1984, p. 191) himself notices, "are likely to be neglected or underemphasized"). It may also lead to disjoint subproblems whose articulation constitutes a problem as such.

*Overestimating the Importance of Search.* Newell and Simon (1972) postulated that "problem solving takes place by search in a problem space" even if this "does not mean that all behavior *relevant* to problem solving" is search. For example, "defining the situation," which may lead to "problem redefinition," "contrasts sharply" with search and thus is "an important type of infor-mation-processing to understand" (p. 761). Neither Newell and Simon (1972), nor Simon (1969/1999) did, however, develop this issue.

"Problem solving as search" models focus on traversal of spaces containing the solutions that are *possible*—in theory (the "basic problem space") or in designers' representation (their "problem space"). The solutions are already *there* and are "simply" to be located in the space in question. In design, problem representations are, however, often not *readily available* through memory retrieval: they have to be *constructed* (cf. design as exploration, Gero, 1998; Logan & Smithers, 1993; Navinchandra, 1991).

*Overestimating the Importance of Means–Ends Analysis.* This main problem-solving method in Newell and Simon's (1972) classical laboratory studies is also supposed to work in design (Simon, 1969/1999, p. 121). In our view, it will, however, seldom be appropriate in design. Indeed, to use this method, one must be "able to represent differences between the desired and the present" (i.e., between the problem's goal state and its current state). For design problems, this will often be difficult. "The" goal state may not yet have been specified but if it has, it may not be done completely or not definitively. Even if the goal and current states have been specified, it will often be difficult, if not impossible, to compare them, as their representations will be at different levels or in different modalities.

*Simon's More Nuanced Positions.* Simon's position with respect to design is not consistent over time. More nuanced positions appear, especially in more recent texts on scientific discovery and creative thinking (Cagan, Kotovsky, & Simon, 2001; Kaplan & Simon, 1990; Klahr & Simon, 2001; Kulkarni & Simon, 1988). Yet, already in his 1973 paper on ill-structured problems, Simon wrote:

> Definiteness of problem structure is largely an illusion that arises when we systematically confound the idealized problem that is presented to an idealized (and unlimitedly powerful) problem solver with the actual problem that is to be attacked by a problem solver with limited (even if large) computational capacities. (p. 186)

More such declarations appear, but they remain isolated in the further context of Simon's work. Simon did not integrate them, for example, in later editions of the *Sciences of the Artificial* (for further details concerning these, perhaps only apparent, contradictory positions, see Visser, 2004).

*A Possible Explanation of Simon's Approach to Design.* Our explanation of Simon's approach to design is based on Simon considering differently

.   •   Cognitive activities in economics and in design.
.   •   Design activities in engineering and in social planning.

Simon's position vis-à-vis cognitive activity displayed in economics seems subtler than with regard to design thinking. Simon is very sensitive to the way in which economic theories idealize human rationality and neglect its limits. He even obtained the Nobel prize in economics thanks to his completely new approach in terms of "bounded rationality," proposed to explain people's economic behaviour—and their satisficing actions. With respect to design, however, Simon surprisingly seems to underestimate human cognitive limitations.

Simon's approach to design seems to be based on his view of what he presents implicitly as the prototype of design: engineering design (and, to a lesser degree, architectural design). Simon states that in social planning "representation problems take on new dimensions" relative to the "relatively well-struc-tured, middle-sized tasks" of engineering and architectural design (1969/1999, p. 141). Simon assumes that for "real-world problems of [the] complexity" of social planning, designers may refer to "weaker" criteria than for standard design. Especially for this type of problems, an appropriate representation may be essential. "Numbers are not the name of this game but rather representational structures that permit functional reasoning, however qualitative it may be." (p. 146).

Thus, Simon's approach of design could be less mysterious if one might suppose that he considers

.   •   *Routine* engineering and architectural design as standard design.
.   •   Social planning as *a special form* of design *radically different* from standard design.

## 2.2. Modern Views on Design. The SIT Approach

From the 1980s onward, researchers have started to formulate, with respect to the SIP approach, alternative views on human activities in their actual accomplishment. In this article we will focus on "*situavity theory (SIT)*," adopting the term proposed by Greeno and Moore (1993) to cover approaches in terms of both "situated cognition" and "situated action." We view SIT as a general framework that may embrace viewpoints on the analysis of design that are more specific. We consider as instances of SIT (e.g., the "reflective practice," "re-flection-in-action," and "knowing-in-action"), on which Schön focuses.

Two groups of studies will be discussed in this section: early SIT studies and research that is more recent.

One other alternative approach to design that deserves to be mentioned separately is the argumentative model proposed by Rittel (1972/1984; see also Rittel & Webber, 1973/1984). "The act of designing consists in making up one's mind in favour of, or against, various positions on each issue" (p. 325).

## Early SIT-Inspired Research

Excepting Rittel, Schön is, as far as we know, the first author after Simon to introduce a new approach to cognitive design theory. His interest in design originated from an educational perspective. He was concerned with the way in which "professionals think in action" as "reflective practitioners" (Schön, 1983) and with the education required to become a practitioner.

Proceeding as an ethnologist–anthropologist in his observations, Schön discusses in detail particular design situations. This leads him to characterise design as a "reflective conversation with the materials of a design situation" (Schön, 1992). Architects (e.g., observing their drawing, transforming it and observing the result) may discover "unintended consequences" of this transformation (Schön & Wiggins, 1992, p. 143). Such "new discoveries" call for new reflection-in-action, resulting in "improvisation on the spot" to respond to the surprises. Even if architects have certain intentions when transforming a drawing, these aims may evolve in "seeing-moving-seeing" cycles (Schön & Wiggins).

For Schön (1988), designing is "not primarily … a form of 'problem solving,' 'information processing,' or 'search,'" but a kind of "*making.*" "Designers make … *representations* of things to be built." "The designer *constructs* the design world within which he/she sets the dimensions of his/her problem space, and invents the moves by which he/she attempts to find solutions" (Schön, 1992, p. 11). Research into problem solving generally considers problems as "given" to the person who is confronted with them. Schön judges that such an approach neglects the process of "problem setting" "by which we define the decision to be made, the ends to be achieved, and the means that may be chosen. In real-world practice, problems … must be constructed from the materials of problematic situations which are puzzling, troubling, and uncertain" (1983, pp. 39–40).

Bucciarelli (1984), an engineer, used participant observation to apply the reflective-practice approach to collective engineering design.

## Current SIT-Inspired Research

It is mainly on a critical basis with respect to Schön—especially in relation to his lack of precision regarding the operationalisation of the notions he pro-posed—that current SIT-inspired research is being developed (Adams, Turns, & Atman, 2003; Dorst, 1997; Dorst & Dijkhuis, 1995; Roozenburg & Dorst, 1999; Valkenburg, 2001; Valkenburg & Dorst, 1998).

Dorst and Dijkhuis compared Simon's and Schön's "fundamentally different paradigms" (Dorst, 1997, p. 204), focusing on Schön's "design-as-experi-enced" view. They consider that "design is not just a process or a profession, "it is *experienced as a situation* that a designer finds him/herself in" (Dorst & Dijkhuis, 1995, p. 264).

Dorst (1997) was one of the first authors to undertake this comparison. In an empirical study, he evaluated the ability of each paradigm to describe "integration," which Dorst considered as "one of the key issues of design-as-ex-perienced." "Someone is designing in an integrated manner when he/she displays a reasoning process building up a network of decisions concerning a topic (part of the problem or solution), while taking account of different contexts" (Roozenburg & Dorst, 1999, p. 34).

Dorst (1997) concluded that the two perspectives are differentially accurate and fruitful for describing and understanding different design phases. Examined from Simon's perspective, a designer works as if design involves "objective interpretation." From Schön's viewpoint, a designer looks at design as if it involves "subjective interpretation." Schön's perspective is most appropriate for conceptual design, less so for preceding phases (Roozenburg & Dorst, 1999, p. 35). For detail design, it will probably also "lose some of its descriptive power" (p. 36). For the information phase, Simon's paradigm performs better. An analysis of the individual Delft protocol (Cross et al., 1996) led Dorst and Dijkhuis (1995) to conclude that it is also appropriate for the embodiment phase.

In an empirical study, Dorst and Dijkhuis (1995) examined how accurately descriptions of design based on Simon's and on Schön's approaches each capture the activity "as experienced by the designers themselves" (p. 261). "Distilling" data-processing systems from both paradigms, the authors use the resulting descriptions to compare their "descriptive value." Considering it essential that a description method preserves the link between process and content in design decisions, the authors evaluate how each system is able to do so.

The results of the rational problem-solving system are presented in four separate graphs. The authors conclude that, using this system, it is difficult to relate process and content. The results of the reflection-in-action system are exemplified in a descriptive summary of the protocol. This allows the authors to narratively link process and content for several central system components. Dorst and Dijkhuis (1995) thus concluded that, in contrast to the rational problem-solving system, the reflection-in-action paradigm makes it possible to relate the design process and its content (p. 274).

Valkenburg and Dorst (1998) investigated the "suitability" of the "mechanism of reflective practice" to describe team designing. Using protocol analysis, they compare two design teams with respect to Schön's four central reflec-tive-practice activities in design: naming, framing, moving, and reflecting. The two design teams vary in their success. Valkenburg and Dorst attributed the disparity to the differences in "effective design time" spent on the four activities. Compared to their unsuccessful colleagues, the successful team spends less time on naming (3% vs. 49%), more time on moving (73% vs. 39%), and reflecting (21% vs. 8%). However, both teams spend the same small amount of time on framing (only some 3% to 4% of their time). One often refers to Schön because of his emphasis on framing in design, but here this activity thus occupies quantitatively a marginal position. Yet, the successful team consecutively develops five different frames, whereas the unsuccessful team develops only one frame.

In a review of several experimental studies conducted on engineering students, Adams et al. (2003) proposed operationalisation of the two notions "problem setting" and "engage in a reflective conversation [with a situation] listening to [the] situation's back-talk." Using verbal-protocol data, the authors compare entering and graduating students. Each of the activities pro-posed as operationalisation of the two notions correlate positively with measures of quality, thus providing substance to these notions.

Compared to their freshmen colleagues, graduating students define more broadly the problem they are solving (a problem-setting attribute), and are more likely to engage in activities that trigger shifts in their mode of analysis, thus accomplishing redefinition of problem-solution elements (effective-lis-tening and reflective-conversation attributes).

Based on their results, the authors conclude that "problem setting and engaging in a reflective conversation across problem setting and problem solving activities are important features of effective design practice" (p. 292).

Valkenburg (2001), in her analysis of Schön's framework, concluded that "not all designing is done in a reflective practice way." In her empirical studies on team design, she also observes "other ways of working": a "wait-and-see attitude" and "[dealing] with the design problem as a set of separate design issues."

Discussion of the SIT Approach

We distinguish the early and more recent SIT-inspired research.
Our critique of early SIT-inspired work concerns

.   •   Its character with respect to theoretical paradigms, which is quite allusive.
.   •   The definition of its central notions, which lacks precision.
.   •   Its methods for data collection, analysis and modelling, which offer no tools to derive higher-level descriptions from the data (see also Nardi, 1996, p. 83).
.   •   The character of its conclusions, which is anecdotal. Early SIT-inspired studies often stop short by merely presenting raw data, rather than providing results likely to be replicable and conclusions likely to be generalisable across situations.

In addition, Schön's ideas on design are based on design students interacting with experienced designers, other than design professionals working for themselves or designing together. They concern "reflection-on-action" rather than reflection-in-action (Valkenburg, 2001).

To conclude: the early SIT-inspired studies present extremely rich descriptions of *unique* designers implementing *unique* activities in *unique* situations.

In more recent SIT-related studies, Dorst and other researchers have been presenting analyses and formulating theoretical frameworks with more precision. Certain authors have started to operationalise the central notions in Schön's work. Their studies constitute examples of ways to directly and indirectly capture and measure these notions in empirical data. Adams et al. (2003)'s work substantiates some of Schön's ideas that had been inspiring, but rather vague in that they could cover extremely diverse behavioural phenomena. Research like theirs opens up a way to develop Schön's framework at a theoretical level and to counter criticisms such as those formulated concerning the early SIT research!

In a comparative analysis of Schön's design-as-experienced view, Dorst and Dijkhuis (1995) concluded that the rational problem-solving system was not appropriate to relate process and content in design decisions. We consider that the different types of description formats used by the authors (separately presented graphs vs. narratively connected descriptions) may have induced this conclusion. One cannot separate the influence of the paradigms from the influence of the description formats.

In spite of several criticisms formulated with respect to SIT—especially Schön's version—we consider the contributions of this approach essential—also thanks to Schön! The insistence on the role of constructive aspects of design, such as modifying design ideas based on new data, but especially adopting new views on "old" data, concurs with our approach to design as the construction of representations.

## 2.3. Confronting SIP and SIT Approaches to Design

Each research community chooses its focus of interest for research—at the cost of simplifying and ignoring other topics. Studies conducted by reference to the SIP paradigm pay more attention to people's *use* of knowledge and representations in problem solving than to the *generation of representations*, and they do not analyse activities as they occur in interaction with other people and the broader environment. SIT-inspired authors focus on the consequences of these *interactions* and the role played by the *environment*, the *social and cultural setting*, and the *situations* in which people find themselves, but they usually ignore thorough analysis, especially of cognitive structures and activities.

We agree with SIT proponents Greeno and Moore (1993) when they contend that, in the SIP theory, "the structure of interactive relations between cognitive agents and the physical systems and other people they interact with" is contained in a black box. SIP theory did not analyse these interactive relations "in anything like the detail that has been characteristic of analyses of hypothesized cognitive structures and procedures."

SIT researchers state that designers' "dominant resource" is the situational context. For SIP researchers this resource seems to be the range of designers' problem-solving methods. Even if the SIP approach does not deny that these resources may evolve under the influence of designers' interaction with their environment, it does not pay much attention to these factors, which play a central role in SIT-related approaches. If SIT approaches do not deny the existence of internal resources, it is only recently that they have begun to analyse them. We saw that the SIT paradigm—especially, in its reflective-practice form—is evolving toward theoretical frameworks that are more substantial.

One might be tempted to conclude with Norman (1993) that the "two traditions do not seem to be contradictory," that "they emphasize different behaviours and different methods of study". Indeed, for a designer confronted with a design task, resources for action—the elaboration of representations, strategies, and other action—depend on both internal (knowledge) and external (social and artefactual) data. We judge, however, that these two types of data are not in a symmetric relation: it is the *designer* who, *using her or his knowledge and representational activities*, establishes the relationship between internal and external data. As these links are to be established by way of representation, we attribute such an essential role to the construction of representations—by the designer!

## 3. DESIGNING AS CONSTRUCTION OF REPRESENTATIONS

Globally characterised, designing consists in specifying an artefact, *given requirements* that indicate—generally neither explicitly, nor completely—one or more functions to be fulfilled, and needs and objectives to be satisfied by the artefact, under certain conditions expressed by constraints. At a cognitive level, this specification activity consists in developing (generating, transforming, and evaluating) representations of the artefact *until they are so concrete, detailed, and precise* that the resulting representation—the *specifications* of the arte-fact—specify explicitly and completely the implementation of the artefact.

The ultimate design representation must express three aspects of the artefact: what—the artefact itself—how—the procedure by which it should be im-plemented—and why—the design rationale (i.e., "the reason why the design should be as it is"; Stacey & Eckert, 2003, p. 170).

The implementation of the specifications into the artefact requires further activities other than designing—even if it is generally still interspersed with more or less important design decisions, because the specifications will not be totally explicit and complete. Indeed, "design never ends." Other design prolongations occur through users' participation in design and through maintenance.

The following two sections discuss the two perspectives that, analytically, can be distinguished in this characterisation of design: the structures that are constructed (evolving design representations) and the corresponding construction activity (its organisation, strategies and activities).

## 3.1. Designs as Representations

There has been much discussion on representations (e.g., about "representation vs. no representation" and "symbolic vs. nonsymbolic representation").

Coming from different origins, Brooks (1991) and Dreyfus (2002) have each, in a paper entitled *Intelligence Without Representation*, claimed that intelligent behaviour does not require representation. In this article, we clearly consider that people construct and use representations—but not of an "original," which would be *re*-presented. *Re-presentation* is a misleading (the German "Vorstellung" is less confusing). It should not be interpreted literally: we do not *re*-present an independent "reality" beyond our experience. This position is close to Von Glasersfeld's (1981) radical constructivism.

Representations are neither "complete" nor "objective." Adapting somewhat Ochanine's (1978) notion, they are "operative" in that they are not as much functionally distorted and restricted to task-relevant characteristics (Ochanine's view), but *shaped* by these characteristics. Variations in cognitive commitment play an important role in this selection.

Representations are also goal-oriented. They are "constructions, which for some purposes, under certain conditions, used by certain people, in certain situations, may be found useful, not true or false" (Bannon, 1995, p. 67). They are built to make things "visible so that they can be seen, talked about, and potentially, manipulated" (Suchman, 1995, p. 63). This visibility is not only for others: an important role in design is played by designers making visible things for themselves, so that, in Schön's terms, they can engage a "conversation" with them, and so advance their design activity.

With respect to the symbolic nature of representations, we refer to the analysis made by Goel in *Sketches of Thought* (1995). Goel is not "against representation", not even "against symbolic representations." Symbolic representations are, however, not all there is.

For Goel (1995) lassical cognitive science is based on what he calls the Computational Theory of Mind (CTM), which claims that cognitive processes are computational and require a representational system with some very stringent properties. These CTM properties are syntactic (disjointness, finite differentiation) and semantic (unambiguity, disjointness, finite differentiation; chapters 7 and 8 *passim*, especially Table 8.1). For well-structured puzzle problems, these properties work well, but in ill-structured, open-ended, real-world problems, such as design, other types of representation also play an essential role. Based on Goodman's (1969, quoted in Goel, 1995) analysis of symbol systems, Goel explains why "sketching" is so essential and typical for design.

Goel's use of the term *sketching* is not restricted to a particular form of pictorial symbol systems. The term refers to the use of any non-notational symbol system (i.e., a system characterised by non-CTM properties), which, in informal terms, may be qualified as imprecise, fluid, amorphous, indeterminate, and ambiguous.

In the rest of this article, "sketch" is used in its common, restricted sense of a particular type of drawing, and not in Goel (1995)'s technical, broad acceptation.

## The Evolving Nature of Representations: Levels of Abstraction

One of the first results in cognitive design studies in architecture was that, in early phases, designers construct and use different representations than in later phases (Goel, 1995; Lebahar, 1983). An important difference concerns two orthogonal abstraction hierarchies:

.     •     The implementation hierarchy (Rasmussen's 1986 abstraction hierarchy): representations differ from abstract, i.e. related to the artefact's purpose, to concrete (i.e., related to the structural or physical properties that specify its implementation).

.     •     The part–whole hierarchy (aggregation hierarchy): representations differ from global to detailed.

We do not adopt the term *refinement* for the second hierarchy, because authors often not only use it for detailing but also for concretisation.

In their activity, designers do not progress through the corresponding representation levels in a systematic, fixed order. Instead, they come and go between representations at level n and level n ± m—even if, globally, there will be more activity at a functional level in early stages and more activity at a structural level in later stages. Goel notices that problem structuring "occurs at the beginning of the task … but may also recur periodically as needed" (1995, p. 114). In addition, "designers differ substantially in the path they take through [the design space] and how quickly or slowly they traverse its various phases" (p. 123; see also McGown, Green, & Rodgers, 1998).

## The Evolving Nature of Representations: Degrees of Precision

A further difference between representations in early and in later phases is that initial representations are necessarily imprecise (Goel, 1995; Lebahar, 1983)—and thus incomplete. During conceptual design, designers need flexible forms of representation that, for one thing, express the provisional character of the underlying ideas, and for another, prevent them from premature commitment to specific options. Only gradually, as design progresses, do representations gain in precision. In domains where external representations play an important role, the type of drawings changes from sketches to drafts with dimensions and tolerances (McGown et al., 1998). Sketches may indeed translate provisionality, a function that in face-to-face, oral communication may be achieved by particular forms of phrasing and intonation (Stacey & Eckert, 2003, p. 167).

External Representations

Recently many cognitive design studies have come to focus on external representations. This interest should not lead to neglecting the importance of internal representations. Even if mental representations have received much attention in traditional cognitive science research, we still lack data on their potential specific features in design. Cognitive design studies discuss "mental models," "mental images," and other internal representations that are supposed to preserve in analogue form, features of the represented entity, but their precise role is not yet clear (Verstijnen, Heylighen, Wagemans, & Neuckermans, 2001; Verstijnen, Leeuwen, Goldschmidt, Hamel, & Hennessey, 1998).

Among the numerous different forms that external representations may take (graphical or verbal–textual; oral or written; two-dimensional or three-dimensional; notes, flowcharts, drawings, plans, or scale models), drawings have received particular attention. Many studies examine their importance, especially in the early phases of design, and particularly in domains concerned with physical artefacts, such as architecture and industrial and mechanical design (Design Studies, 1998; Do, Gross, Neiman, & Zimring, 2000; Goel, 1995; Kavakli, Scrivener, & Ball, 1998; McGown et al., 1998; Neiman, Gross, & Do, 1999; Rodgers, Green, & McGown, 2000; Scrivener, 1997; Tseng, Scrivener, & Ball, 2002; Verstijnen et al., 2001; Verstijnen et al., 1998). In these domains, "visual representations are omnipresent throughout the design process, from early sketches to CAD-rendered general arrangement drawings" (McGown et al., 1998). According to Hwang and Ullman (quoted in McGown et al., 1998, p. 432), "67% of the marks made on paper during conceptual activity are sketches."

An interesting characteristic of drawings is that, unlike oral expressions, they leave traces—even if many exploratory drawings are thrown away (McGown et al., 1998). Later on, designers can come back to these residual representations, examine them at ease, show them to colleagues (cf. also their possible role in maintenance; e.g., in forms of design rationale).

Important functions of external representations depend on the externalisation and visualisation they allow and that may facilitate designing. They enable activities on the entity represented—generally, the artefact under design—that are more difficult or even impossible on the corresponding internal representations. Using an external representation, it is often easier to "manipulate" an entity, to reason on it, to test hypotheses and other ideas on its subject. External representations often facilitate the exploration of alternatives, the prediction of outcomes or consequences of new ideas (Do et al., 2000). Through the possibilities of simulation, drawings may serve to generate solutions: it is easy to try out quickly, and cheaply, different options. They are also useful in evaluation: juxtaposing various drawings or mock-ups, a designer may compare different possibilities. These instrumental functions of external representations are essential for designers, to advance their design, work on it through controlled reasoning activities, but also unintentional discovery. Intermediary results of designers' activity in the form of external representations, may lead them, in "reflective conversation" with these representations, to evolve in their interpretations, intentions and ideas for solutions.

External representations are also of course helpful as memory aids in an extension of internal memory (to temporarily stock provisional ideas, and permanently archive intermediate or final solutions).

"Computational offloading" refers to the observation that, compared to internal representations, the use of external representations reduces the amount of cognitive effort required to solve informationally equivalent problems (Larkin & Simon, 1987; Zhang, 1997).

*Sketches.* These preliminary drawings for later elaboration, are generally drawn by hand; few computerised systems really allow sketching (but see Decortis, Leclercq, Boulanger, & Safin, 2004; Leclercq, 1999).

Thanks to their fluidity and imprecision, sketches are supposed to give access to knowledge not yet retrieved, to evoke new ways of looking and seeing and, thus, to result in an enhancement of creativity and innovation in design. In research in visual, graphic brainstorming, Van der Lugt (2000) has analysed this claim. Given his conclusion that "in early idea generation, sentential, or partly sentential, variations of the brainstorming tool … perform stronger than [their] graphic variations" (p. 521), one may claim that external, graphical representations are not the panacea for creativity enhancement!

Interdesigner Compatible Representations in Collective Design ("Shared" Representations)

Given that they associate components from various domains of specialty, design projects generally require multiple skills—and thus collaboration.

The role of representations in collective design varies according to its phases. During distributed design, designers each have their own tasks and specific goals to pursue. When codesigning, they have a common goal that they aim to reach by applying their specific skills and expertise. It is then essential that designers who each also have their personal perspective,5 establish a "common ground" (Clark & Brennan, 1991) or a "common frame of reference" (De Terssac & Chabaud, 1990). These representations concern agreements, especially on the definition of tasks, states of the design, references of central notions, and weights of criteria and constraints. They are often qualified as "shared," but given the fact that one has no objective reference, we prefer to characterise them as inter-designer compatible representations (cf. Von Glasersfeld, 1981).

Representations at the Source of a Design Project: "Design Problems"

The representation at the source of a design project expresses the requirements for the artefact—or, in problem-solving terms, the problem to be solved.

Qualifying design as a "problem" refers to designers' representation of the task that consists in satisfying the requirements. It does not convey any presupposition with respect, either to the requirements being exhaustively and unambiguously developed when the designers receive them (Carroll, Rosson, Chin, & Koenemann, 1997, 1998), or to a definitive representation of the requirements being constructed by the designers when they first analyse them.

---

5 In recent years, the notion "viewpoint" has been proposed in many publications, without having received a clear definition that distinguishes it from other representations. We use the term *perspective*, not in a technical sense (i.e., to refer to a particular type of representation), but to focus on the fact that different designers may have constructed different representations of a "same" entity (artefact or other), and that in their representation of a "same" entity, different aspects (attributes) may receive different weights.

The "design as a problem" approach has been the object of many SIP-in-spired cognitive design research studies (see references in Bayazit, 2004; Cross, 2001, 2004; Détienne, 2002; Eastman, 2001). In this section, we will therefore discuss only two characteristics on which we propose a complementary or different view. These are design problems' ill-definedness and the way in which the artefact is constrained by these representations that constitute design problems.

*Ill-Definedness.* Instead of *ill-structured* we adopt the term *ill-defined* to refer to both problem's ill-structuredness, and the ill-specified character of all three problem components: initial state, goal state, and operators. This position is inspired by Reitman (1964; see also Thomas & Carroll, 1979/1984). Many authors consider design problems to be ill-defined because of one or two components being ill-specified. They, for example, propose that de-sign-problem specifications have open constraints, or that design problems admit several solutions.

Reitman (1964) distinguished different types of problems according to the degree of specification of each component. In design problems, only the goal state generally receives some explicit specification, by a description of the artefact's function and of other constraints on the artefact—even if this specification is abstract and incomplete. Initial state and operators are generally underspecified. Implicitly, they may be supposed to correspond to the state of the art in the domain.

Problem-definedness is clearly a relative characteristic (cf. also Newell, 1969, p. 375; Schön, 1988, p. 184).

*Constraints on the artefact.* Several authors, especially in A.I., analyse design as constraint management and satisfaction (Logan & Smithers, 1993; Stefik, 1981a, 1981b). Darses (1990a; 1990b), analysing such approaches from a cognitive–ergonomics viewpoint, concludes that, even if these interdependent variables play a central role in design, designers' activity also refers to other sources for action (e.g., action plans and schemata).

The constraints that design problems are to satisfy are often conflicting. Dealing with trade-offs among constraints, and with different possibilities to do so, is central in satisficing. Combined with design's complexity (i.e., its multidimensional and interdependent organisation), this leads to making the analysis, structuring, and pruning of constraints into an essential aspect of design activity.

Even if, formally, it is generally possible to transform an ill-defined problem into a well-defined one by closing all open constraints (Reitman, 1964), such a transformation corresponds to a form of premature commitment that designers may well regret afterwards.

According to a common-sense belief, constraints are often considered as obstructing people's "freedom." However, constraints are useful, if not necessary, in narrowing down a space of possibilities that may otherwise be too large for exploration. "Even" the expression of artistic creativity requires constraints (cf. Stacey & Eckert, 2003, p. 166).

Using Knowledge in Design

The importance of knowledge holds for most professional domains, but it is particularly critical in an activity that essentially consists in representational activities. Design requires general, abstract knowledge and weak, generally applicable methods, but above all designers need domain-specific knowledge, and the corresponding strong, knowledge-intensive methods. We suppose that satisficing (e.g., requires more domain-specific knowledge than does optimising). This also holds for the creativity that is particularly important in design compared to routine activities. In addition, knowledge is a key element in the exercise of analogical reasoning—which may, in turn, be related to design creativity (but see Visser, 1996).

Nonalgorithmic, nonformalised activities—necessary in creativity, satisficing, (re)interpretation, and qualitative simulation—require knowledge. Of course, to proceed to complex calculations, a designer also needs knowledge, but of a sort that can be learned. The knowledge that is very important in design is not gained through formal education, but through experience. Designers may acquire such knowledge as a result of their work on many different types of projects, and their interaction with colleagues who have other specialties.

Without knowledge, no interpretation, thus neither the possibility to look at a project in a way different from one's colleagues, nor that of seeing things differently than one did during a previous project! The task-and goal-ori-ented character of representation results from an interaction between one's knowledge and experience, and the situation one is in.

Knowledge determines if a design task constitutes a problem. Working with ill-defined problem data is only possible if one has specific knowledge—in addition to generic knowledge.

Furthermore, knowledge is a critical resource underlying most strategies. If simulation via representations works, it is thanks to one's knowledge. Reuse is, by definition, impossible without knowledge (it is not a components library that makes knowledge superfluous). Handling constraints (especially constructed constraints) would be hard without it.

These are only a few examples, mentioned to indicate the importance of knowledge in design. Several related issues are not discussed here: distinctions between knowledge in the application domain, underlying theoretical and technical domains, and relevant nontechnical domains (Visser, 1995b), knowledge concerning design methods, and knowledge of ergonomics, and social, political, economic, or legal aspects of the artefact and its use. As designers generally are not expert in all these different domains, the need of design projects for wide-ranging knowledge requires collaboration between professionals from various domains.

## 3.2. Designing as an Activity: Construction of Representations

After a short discussion of the relations between designing and the tasks surrounding it, we discuss consecutively the activity at three levels: its organisation, its strategies, and its activities.

### The Intermingled Character of Design and Its Surrounding Activities

Requirements specification, the task that precedes design, is not supposed to involve designing. According to a common viewpoint, requirements are "out there" and can fully be gathered at the beginning of a design project. In the course of a series of participatory design sessions, Carroll et al. (1997, 1998) noticed, however, how project requirements evolved. "The client's original functional requirements … were radically and continuously transformed … Qualitatively different requirements become accessible or salient" (1998, p. 1167). The authors did not analyse this as "initially mistaken notions being subsequently corrected," or as "more requirements work leading to successively finer decompositions" (1997, p. 62). They qualified the process of collaborative requirements gathering as requirements *development*—in our terms, requirements *design*.

A comparable relation may be identified between design and implementation. Generally there will be, during the task that follows design in an arte-fact's development, decisions that, from a cognitive viewpoint, may be qualified as design. Correspondingly, already during design there are implementation activities. In our software-design study, we observed that specifying software and not coding it right away may be difficult for software designers (Visser, 1987).

Organising One's Design

The interwoven character of activities also holds at a lower level, inside the global design task. The process followed by designers in industrial, complex projects does not progress through independent consecutive stages. Designers do not first structure "the" problem and then solve it: design is not a process going from analysing "the" problem specifications to synthesising "the" solution. Neither do designers systematically traverse the three stages often distinguished in problem solving: construction of a problem representation, solution generation, and solution evaluation.

Since the 1980s, an increasing number of authors have claimed, based on empirical design studies, that the actual organisation of design activity is not a simple implementation of systematic decomposition strategies (especially, top-down combined with breadth-first or depth-first), following preestablished plans (especially, hierarchical plans). The qualification that has come to qualify the way in which designers organise their activity is "opportunistic" (Visser, 1987, 1994) or "serendipitous" (Guindon, Krasner, & Curtis, 1987).

Like many colleagues (see, e.g., Ball & Ormerod, 1995), we assume that designers have principles that guide their activity. Designers are aware that the use of a combined top-down breadth-first strategy is a valuable approach to organise their design properly (e.g., to obtain well-structured specifications, think of all design components, and correctly handle their interactions). They are conscious that the avoidance of premature commitment is precious and that the breadth-first decomposition minimises this risk. However, designers, for one thing, meet difficulties in the implementation of such systematic strategies, and, for another, use other resources to organise their activity than simply the structured plans provided by systematic strategies. These strategies indeed impose a heavy load on memory and hinder exploitation of possibilities for action that may be interesting for different reasons.

To discuss these issues, we need to anticipate the general presentation of design strategies, and present these decomposition strategies.

*Top-Down and Bottom-Up Strategies, Breadth-First and Depth-First Strategies.* The top-down strategy consists in descending a problem's theoretical "solution tree" from the most abstract level down to the lowest, concrete and detailed level, without ever returning to a higher level.

In addition, design methodologies advocate that one combine the top-down strategy with a breadth-first strategy: when decomposing a prob-lem-solution, one should develop, consecutively, all the elements of the current problem-solution at the same level of the solution tree, before going on to the next level.

Several authors of early empirical studies conducted on design, especially in the domain of software design, claimed that the integrated top-down, breadth-first decomposition, was usually indeed the global strategy implemented by experts (Adelson, Littman, Ehrlich, Black, & Soloway, 1985; Adelson & Soloway, 1988; Ball & Ormerod, 1995; Byrne, 1977; Davies, 1991; Jeffries, Turner, Polson, & Atwood, 1981).

However, top-down balanced refinement seldom occurs absolutely. Some 10 years ago, we analysed the results of 15 empirical studies on design in diverse domains, focusing on designers' organisation of their activity (Visser, 1992, 1994). We concluded that, with one exception (Adelson & Soloway, 1988), each study showed one or more factors that contributed to the opportunistic character of the organisation of design.

*The Opportunistic Organisation of Design.* Hayes-Roth and Hayes-Roth's (1979) study on errand planning is generally *the* reference for qualifying design organisation as opportunistic. We have applied the authors' view of opportunism to more "typical" design tasks (functional specification, Visser, 1994, and software design, Visser, 1987) (see also Bisseret et al., 1988, for traf-fic-signal setting, and Whitefield, 1986, for mechanical design). Our focus was designers' organisation of their activity (Visser, 1990).

*Opportunism* is not just a "blanket-term to denote *any* design activity that deviates from a single, rigid design approach" as Ball & Ormerod (1995) considered opportunism proponents to do. We attribute the particular nonsystematic character that we qualify as opportunistic, to the fact that designers, rather than systematically implementing a structured decomposition strategy, take into consideration the data which they have at the time: specifically, their knowledge, the state of their design in progress, their representation of this design, and the information at their disposal (Visser, 1994). Considering this data in addition to the possibilities provided by systematic decomposition, designers considerably increase their potential range of action.

In our studies, we identified six categories of data as being factors that lead to the opportunistic organisation of design. They range from information provided to the designer by an external information source (in particular, the client or a colleague), via information the designer "comes across" when "drifting" (involuntary attention switching to a design object or action other than the current one), to mental representations that are activated by the representations used for the current design action (e.g., because of analogical relations between both representations) and which may concern design objects or procedures (for more details, see Visser, 1990, 1994).

*Discussion of Our Opportunistic-Organisation Position.* Not all researchers share the view that design is organised opportunistically. Davies (1991) has characterised program design as "broadly top-down with opportunistic local episodes" (p. 176)—exactly the opposite of our "organisation of de-sign activities [as] opportunistic, with hierarchical episodes" (title of Visser, 1994)! With respect to this issue, we follow Hayes-Roth and Hayes-Roth (1979, p. 307) who, to solve "the apparent conflict" between the two models, proposed considering the top-down, successive refinement model as a special case of the opportunistic model. Indeed, the model that allows for various or-ganisational structures is more general than a model that allows only for one or two structures in articulation (top-down combined with breadth-first or depth-first). Yet, we favour an opportunistic model not only because of this formal argument but also for "positive" reasons (see following).

With respect to these modelling choices, several questions may be raised. How great a lack of systematicity can be tolerated by a systematic-decomposi-tion model? Alternatively, what degree of systematicity can be qualified as a particular form of "opportunism," before this qualification becomes inappropriate?

In his conclusion, Davies (1991) noted that the differences between experts and novices may be due to differential "use and reliance upon external memory sources" (p. 186). This point becomes central in Ball and Ormerod's (1995) discussion. These authors argued that "much of what has been described as opportunistic design behaviour appears to reflect a mix of breath-first and depth-first modes of solution development" (p. 131), even if design is also "subject to potentially diverging influences such as serendipitous events and design failures" (p. 145).

Our view on these positions is that we do not deny that designers may proceed in a breadth-first or depth-first way, but we want to emphasise that

.       •       They often do so *occasionally* and *locally*, rather than *systematically* and *throughout their design process.*

.       •       A breadth-first–depth-first mix can take different forms, and *if* a mix pattern has various occurrences, these will generally be interspersed with other ways of proceeding, involving the risk that the breadth-*first* and depth-*first* qualifications loose their applicability.

.       •       An occasional, local breadth "first"–depth "first" mix is just one form in which opportunism can reveal itself in design.

If design was in effect "simply" "a mix of breadth-first and depth-first modes," it might indeed be exaggerated to resort to opportunism that, in this case, opens up a needless large space of possible forms of structuredness and unstructuredness.

With respect to the structured character of design organisation, opportunism proponents (Guindon et al., 1987; Kant, 1985; Ullman, Dietterich, & Staufer, 1988; Visser, 1988; Voss, Greene, Post, & Penner, 1983) do question the systematic implementation not only of breath-first and depth-first approaches but also of systematic top-down refinement, an issue that Ball and Ormerod (1995) omit completely from their discussion. However, Jeffries et al. 's (1981) results, which are one of the foundations of Ball and Ormerod's (1995) claim, show that the top-down strategy also is not systematically implemented by expert designers.

We close this section by a short discussion of the "potential causes of unstructured activity" (Ball & Ormerod, 1995, p. 145). Ball and Ormerod (1995) conceded that "like Guindon (1990) and Visser (1990) [they] recognize a number of [such causes]. These include factors such as social influences, memory failures, design failures, information unavailability, boredom and serendipitous events" (p. 145).

We suppose that most, if not all, cognitive functioning is, for an important part, under the dependence of people's limited information processing capacities (cf. Simon's bounded rationality). We suppose, however, that in addition to the more or less frequently arising *impossibility* of working in a systematic way because of cognitive *deficiencies*, there are also positive causes of opportunism. Designers also *decide* to take *advantage* of opportunities, in spite of other possibilities. The decision to exploit analogical relations, information available because a colleague passes by, or "discoveries" that emerge as by-products of systematic design actions, are all examples of instances of such positive factors leading to an opportunistically organised activity.

Design Strategies

Given the focus of this article, we will discuss strategies that are used by experienced, professional designers (rather than students, as in most studies on this topic; see also Akin, 1979/1984; Gilmore, 1990; Visser & Hoc, 1990).

*Reuse Versus Design From Scratch.* All use of knowledge could be qualified "reuse" in that knowledge is based on the processing of previously encountered data, experience, and representations constructed and used in the past. We have proposed to reserve "reuse" for the use of specific knowledge that is at the same abstraction level as the target for whose processing this knowledge is retrieved (Visser, 1995a; cf. case-based reasoning, Visser & Trousse, 1993).

Reuse has been observed in various empirical design studies. The exploitation of specific experiences from the past is indeed particularly useful in design, especially in nonroutine design (which might seem surprising, but see Visser, 1996).

Many aspects of reuse have been examined in cognitive design studies: its phases, strategies, types of reusable entities and entities reused, types of exploitation, effects of reuse on productivity, and difficulties and risks of reuse (see Détienne, 2002; Visser, 1995a). In this article, we only hint at the question concerning the origin of reuse in design. Under which conditions do designers decide to solve a problem by reuse, rather than refer to general knowledge (i.e., design from scratch)? Designers need to take this decision during the construction of a target-problem representation. Both Burkhardt and Détienne (1995) and Visser (1987) concluded that the cost of reuse is the main factor in this process (an observation in line with the importance of cognitive economy as primordial criterion underlying design organisation (Visser, 1994).

*Simulation.* Simulation, often presented as an important evaluation strategy, is also used in generation. Mental simulation may serve to understand and elaborate the problem requirements (Guindon et al., 1987; Kant, 1985; Ullman et al., 1988) or to evoke from memory knowledge relating to a problem (Baykan, 1996, p. 141).

In design domains involving physical artefacts, designers also simulate by means of external representations such as drawings and mock-ups. In a comparative analysis of manual and computer-assisted design, Dillon and Sweeney observed that such simulation "acted as a powerful visual aid" to some of the manually working designers, whereas "no directly comparable facility existed on CAD" (1988, p. 484).

*Selecting a Kernel Idea and Premature Commitment.* Several authors have observed that designers, early on in their activity, tend to select a "primary generator", that is a few simple objectives adopted to generate the initial solution kernel (Darke, 1979/1984). They then stick to such a "central concept" in what is going to become their global design solution (Guindon et al., 1987; Kant, 1985; Ullman et al., 1988).

One might thus conclude that designers neglect to elaborate and compare alternative representations of design problem-solutions. The literature provides, however, other observations. In a study on professional design review meetings, Ball and Ormerod (2000) conjectured that it is individual or de-contextualised design that induces such an early fixation. They indeed observe that the design team explores a range of alternative concepts. In our study on industrial composite-structure design by a team of mechanical designers (Visser, 1993), the two most experienced designers were observed to come up with a multitude of solution principles. There are nevertheless also experimental studies of individually conducted design in which designers formulate several solution ideas (e.g., Eastman, 1969).

Selecting a kernel idea early on in the design process and sticking to it, may be considered a form of premature commitment. Several authors noticed that designers often try to avoid such early fixation (Lebahar, 1983; Reitman, 1964). The apparent contradiction may be removed in at least two ways:

.  •  Designers may declare or aspire to refrain from premature commitment, but in fact not put these intentions in practice (cf. our observation that designers' accounts about their activity often do not coincide with their actual activity, Visser, 1990).

.  •  Early on in the design process, designers may select a kernel idea at a conceptual level, but later on, they may refrain from premature commitment at a more concrete or detailed level, by not fixing all values for its variables.

Design-Representation Construction Activities

Many recent studies concern representational structures in design, especially external representations, but the cognitive activities and operations[6] involved in their construction and use have not been analysed extensively.

We distinguish three types of activities on representations: generation, transformation, and evaluation. We also discuss the use of representations in collective design.

*Generation and Transformation of Representations.* The first two "stages" distinguished from a problem-solving perspective (construction of a problem representation, and solution generation) both involve construction of representations, even if the input and output representations are different.

A representation is never generated "out of nothing." One always transforms representations, from the first design action onward. Yet, we qualify the construction of representations as "generation" if one's memory is its main source. It is never its only source: both the state of the design project (requirements and their follow-up included) and other contributions "from the outside world" will influence a designer. Designers will usually generate a first representation of a design project by interpreting the requirements.

Generation may be implemented by different types of processes and operations: from the simple evocation of knowledge from memory, to the elaboration of "new" representations out of mnesic knowledge entities without a clear link to the current task (Visser, 1991). Schema instantiation is the form of knowledge evocation that has received much attention in software-design studies (Détienne, 2002).

---

[6] Except when an action is clearly an operation, we will use the term *activity* to refer to activities and operations.

We distinguish transformation activities according to the type of transformation between input representation $r_x$ and output representation $r_y$. They may *replicate* (reformulate $r_i$ into $r_i$', or duplicate it, Goel, 1995), *add* (transform $r_i$ into $r_i$' by introduction of "small alterations", Van der Lugt, 2002), *detail* (break up $r_i$ into components $r_{i1}$ to $r_{in}$), *concretise* (transform $r_i$ into $r_i$' that represents $r_i$ from a more concrete perspective), *modify* (transform $r_i$ into another version, $r_i$', neither detailing, nor concretising it), or *substitute* (replace $r_i$ by an alternative representation $r_j$, neither detailing, nor concretising it; cf. Van der Lugt's 2002 "tangential transformations" (i.e., "wild leaps into a different direction").

Many activities play a more or less direct role in these different types of transformation. Some examples (varying between operations and activities) are interpretation, association, brainstorming, reinterpretation, confrontation, adjustment, integration, analysis, exploration, inference, restructuring, combining, drawing (sketching, drafting, and other forms), hypothesising, justifying. In this article, we only comment on some of them.

We do not present activities that correspond to configurations of other activities, such as cognitive synchronisation, or conflict resolution, which we analysed in terms of "exchanges" (polylogal verbal–interactional units) in our analysis of technical-review meetings (D'Astous, Détienne, Visser, & Robillard, 2004). Neither do we discuss basic processes (mechanisms). Finally, we do not include activities at the action-management level, such as planning, organisation, regulation, or control.

Even if it is too simplistic to describe a first design "stage" as "analysis," it corresponds to a central activity in the initial phases of a design project. Constraints analysis is essential to disambiguate design requirements. Analysing the current design state may be a way to introduce detail or concreteness in the project.

Analogical reasoning occurs in all three representational reasoning activities. We have mentioned it in different contexts: as a factor of opportunism, in creativ-ity-requiring activities, to tackle ill-defined problems by interpreting them, and to generate "interesting" design ideas.

Different forms of inference are used in design. Induction is much more frequent than deduction. Goel (1995) identifies only 1.3% "(overt) deductive inferences" in his observations. In our composite-structure design study (Visser, 1991), neither did we notice any overt form of deduction.

Restructuring and combining representations are often mentioned as components of the creative process (Verstijnen et al., 2001; Verstijnen et al., 1998). Verstijnen et al. show that they are two separate constituents of creativity that function differently. Restructuring an existing external representation (drawing) is difficult to perform mentally, and is facilitated—but only for experienced designers—if one is allowed or encouraged to sketch. Combining (synthesising) parts of such a representation, however, can be performed by only using mental imagery.

As tools for reinterpretation, activities such as restructuring and combining may be used to generate new ideas. Drawing is also a tool for other activities. It may serve restructuring, combining, analysis, or simulation. It may also fulfil interactional functions, such as informing, or explaining. It can even have several functions simultaneously (e.g., simulation, explanation, and storing).

Confrontation, adjustment, and integration of representations play a specific role in collective design.

*Evaluation of Representations.* According to design methodologies, the generation and evaluation of solutions are two different stages in a design project. Many empirical studies have shown, however, that designers intertwine the two. The participants in the technical review meetings that we studied (D'Astous et al., 2004) were supposed to follow a particular method in which design was not supposed to occur. They came up, however, with alternative solutions (i.e., they were not only recording the underlying negative evaluations).

Evaluating an entity consists in assessing it vis-à-vis one or more references (Bonnardel, 1991). In the context of design, evaluation may occur when a representation is presented by its author, or interpreted by colleagues, as an "idea" or "solution proposal." Colleagues may interpret a representation as a solution proposal without its author presenting it explicitly as such, and they may evaluate it without its author explicitly requesting them to do so (Visser, 1993).

According to their source, different types of evaluative references are distinguished (Bonnardel, 1991; Ullman et al., 1988), and depending on the type of reference used, evaluation may adopt different strategies (Martin, Détienne, & Lavigne, 2000, 2001).

Evaluative references are forms of knowledge. As expected, designers' expertise in a domain influences their use of these references (D'Astous et al., 2004).

Given that in a collective design setting, designers may have different perspectives regarding a project, proposals are not only evaluated on purely technical, "objective" evaluative criteria. They are also the object of negotiation, and the final agreement concerning a solution also results from compromises between designers (Martin et al., 2000, 2001). In addition, not only solution proposals, but also evaluation criteria and procedures undergo evaluation (D'Astous et al., 2004).

Evaluation has functions at both the action–execution, and the ac-tion–management level of the activity. Designers do not only evaluate design objects (solutions, criteria), but also their design process, its progression and direction (Visser, 1996).

*Construction of Interdesigner Compatible Representations.* We consider that the main function of both language[7] and drawing, in collective design, is not the simple expression of ideas previously developed in an internal medium. Besides the functions that representations play in both individual *and* collective design settings, various aspects of the externalisation possibilities of representations provide additional functions specific to collective design. These functions go together with different cooperative activities, which vary according to the phases of the design project.

During distributed design, where the designers' central activity is coordination to manage task interdependencies, representations of course play a role. Yet, it is in codesign that they have a particular function due to its collective setting.

Construction of compatible representations when co-designing proceeds through activities qualified as "grounding" (Clark & Brennan, 1991) and "cognitive synchronisation" (D'Astous et al., 2004; Falzon, 1994). D'Astous et al. (2004) showed that constructing such representations of the to-be-reviewed design solution was a prerequisite for the prescribed task to occur (i.e., evalua-tion of this solution).

Given that designers have their personal perspectives, collaboration between designers calls for confrontation, adjustment, and integration of these different representations for the designers to be able to reach a common solution. A great amount of time is spent on these activities (D'Astous et al., 2004; Herbsleb et al., 1995; Karsenty, 1991; Olson, Olson, Carter, & Storrosten, 1992; Stempfle & Badke-Schaub, 2002).

---

[7] In this article, "language" only refers to verbal language, not to other semiotic systems.

The specific benefit of visual expression in creative collective activity has been examined by Van der Lugt (2002). He showed that sketching using brainsketching tools contributes to creative activity in idea-generation groups, but not as expected: it especially supports reinterpretation of one's own ideas, and so stimulates creativity in individual idea generation. Reinterpretation of ideas generated by other group members is not enhanced. Collective working is thus not the panacea for all complex processes.

Another interactional situation is the communication between designers and users. With respect to interactive-software design, Carroll (2002) noticed that there is a big and crucial "gap" between the worldviews held by designers of software and its potential users. Participatory design is one way to bridge this gap.

Argumentation—a "hot item" in studies on cooperative activities—has only been touched upon in this article (cf. Rittel's 1972/1984 argumentative model). Authors attribute a more or less broad sense to the notion. We conceive argumentation as an attempt to modify the representations held by one's interlocutors. Many activities in codesign are thus argumentative.

In collective design, representations may function as "boundary objects" (Star, 1988), which serve as an interface between people from different "communities of practice." They can take on many different forms, both representational and artefactual. Their succeeding does not mean that people view them in the same way.

An example of "boundary *representations*"(the term is ours) are the "technical sketches" that the knitwear designers examined by Eckert (see Stacey & Eckert, 2003, p. 163) use to communicate with the technicians who are to realise the garments. They do not work as boundary objects, because they do not contain sufficient detail to be understandable by the different parties involved.

## 3.3. Discussion

Our approach to design will be discussed in the general Conclusion. Here we restrict ourselves to some specific points to which we will not return. Given the focus of this journal, design of HCI receives special discussion. We further discuss the rare cognitive design research that has been concerned with the result of design. A third discussion point concerns the assumption that certain dimensions of the object of design influence the design activity,
i.e. the idea that, in spite of generic design characteristics, there are different "forms" of design. Finally, we prolong the discussion of representational structures in design and the activities operating on them. The next—and necessary (Anderson, 1978)—step is to link the two in a process model of design. Based on tendencies identified in cognitive design studies, we will outline here some initial directions regarding functional linkages between representational structures and activities.

### The Potential Specificity of HCI Design

Design in the domain of HCI has been widely discussed for some 25 years. Some authors notice the similarities between the design of user interfaces and that of other types of artefacts (e.g., architecture or automotive design). In all these domains, "a good design relies on principles such as ease-of-use and providing functionality that meets real needs" (Ford & Marchak, 1997). Winograd (1996) considers that its user-oriented character makes software design comparable to architectural and graphic design, and different from engineering design. He also considers, however, that the design of interactive software is completely different from other software design (Winograd, 1997). Among the arguments advanced for all these claims, none are based on cognitive analyses of the activities.

In the literature that seems to tackle design of HCI, design *activity* is often "discovered" as an unexplored continent. Presenting some observations, evoking one or more questions, authors advance some general ideas without any reference to the numerous results and models in the domain of cognitive design research. This approach may even be found among those involved in CHI, *the* conference in the domain of HCI. Both empirical work and modelling approaches of the cognitive aspects of design activity are rare among contributions to this conference. In 1995, CHI "instituted a new section called design briefings for presentation of notable designs and for discussion of how those designs came to be" (Winograd, 1996, p. xiv). These contributions did not discuss, however, the cognitive aspects of this "coming to be."

Another example is a paper in the Special Interest Community in Com-puter-Human Interaction (SIGCHI ) bulletin. Dykstra-Erickson, Mackay, and Arnowitz (2001) proposed a *starting point* for the discussion of design in SIGCHI (p. 109; emphasis ours). Seemingly unaware of the existing tradition in cognitive design studies, the "fundamental topic" of the paper is the question: "what is 'design,' and how do we define 'designers'?" (p. 109). To answer these questions, the authors introduce three different "perspectives on design": Norman's culturo-organisational view, Sarah Kuhn's sociotechnical approach, and Kelley's idea that "design" and "designer" have a much broader coverage than is generally acknowledged and that design, amongst other aspects, comprises visualising." A dilemma for the CHI community [is that] 'design' really isn't something that can be narrowly defined" (p. 113). The authors "propose" that design "needs to be further qualified" and that this could take place "at the CHI table" (p. 113).

Many studies in the domain of "interactive computing systems for human use" (as ACM SIGCHI defines HCI) seem to be related to designing. Nevertheless, little attention is paid to the cognitive, dynamic aspects of the underlying activity (our software-design study has disclosed some aspects of the way in which a designer's activity may be guided by user considerations; Visser, 1987).

Studies on design knowledge considered relevant for HCI designers rarely analyse *the way in which* designers *use* such knowledge. Referring to authorities such as Norman and Draper (1986), authors advance that the HCI-design model emphasises user-centredness during the design process, but they do not describe *by which cognitive activities* designers of HCI systems may *realise* this user-centredness in their activity.

In theory, the design of HCI might exploit certain conclusions from soft-ware-design studies, which historically occupies a strong position in cognitive design research. However, many of these studies provide little data on actual design. First, most of them concern "programming" in the sense of coding (i.e., implementation of design decisions, rather than the elaboration of these decisions; cf. the designation of "Psychology of programming" and the corresponding Empirical Studies of Programmers workshops; see also Gilmore, Winder, & Détienne, 1994; Hoc, Green,

Samurçay, & Gilmore, 1990). Second, most studies concern small tasks—be it design or coding—examined in artificially restricted laboratory conditions, whereas several critical, distinctive characteristics of design only appear in real, professional design. In our software-design study, we corroborated the hypothesis that strategies used in a professional design setting differ, at least in part, from those used by novice, student programmers working on limited problems (Visser, 1987). Recent software-design studies focus on teamwork, but the collaboration concerns designers, not designers and users.

The absence of studies in the domain of HCI design on cognitive, dynamic aspects of the activity may seem surprising. It may be due to the scarcity of cognitive psychologists in the HCI community. Ergonomists and others working in the domain of human factors, cognitively oriented or not, indeed often lack a research tradition into cognitive activities—at a microscopic level, at least (Garrigou & Visser, 1998).

This discussion is not intended to suggest that the research in HCI is not occupied with relevant cognitive aspects. The artefact's impact on the user and its consequences for user involvement in design are research topics that are central to this domain and that have not been much examined in other domains of design.

Yet, at the end of this section, the question remains: Is there specificity to HCI design—and so, what is it? (cf. our section on "Different Forms of Design").

Design Quality

It seems significant to us that cognitive design research has not been greatly occupied with the result of design. Researchers in this domain have noticed that an artefact may take different forms without one being "better" than the other, but they have rarely examined cognitive factors underlying design quality. This quality remains mostly the affair of design engineers and methodologists, who are concerned with measuring, or estimating, both the effort put into the process and the quality of the result (see Jedlitschka & Ciolkowski, 2004, for software). We have defended elsewhere the idea that measuring process effort and product quality, and establishing a relation between the two cannot be performed without a model of the cognitive activities involved in the design task, and without a measurement of these activities (Détienne, Burkhardt, & Visser, 2004). Today, the data that cognitive models may provide regarding this issue is, however, still sporadic. It is probably no coincidence that the only cognitive design study on this theme has been conducted in the framework of a collaboration project between psychologists and engineering-design researchers. In this project, Fricke (1999) identified several characteristics of the activities of successful designers. Cognitive design studies, however, have shown at length that designers often do not implement the corresponding successful procedures or strategies.

Different Forms of Design

"Activities as diverse as software design, architectural design, naming and letter-writing appear to have much in common" (Thomas & Carroll, 1979/1984). There are, however, not only important similarities between design activities in different domains, but also important differences between design tasks and nondesign tasks (Goel, 1994; Thomas & Carroll, 1979/1984). We adhere to this idea of "generic design," and yet, we suppose that the nature of the artefact introduces specificities in the corresponding design activities.

As far as we know, this assumption has never been the object of specific cognitive analysis. Without presenting any empirical or theoretical evidence, Hubka and Eder (1987) advanced that the type of design object influences the activity.

Our discussion starts, once again, with Simon's position. If Simon adhered to the hypothesis we are formulating here, this could explain, at least in part, his different views on standard design and social planning. His publications, however, did not discuss this idea.

In this section, we will propose some candidates for dimensions underlying differences between forms of design.

We do not deal with the effect that the use of different design methodologies may have on designing and on the resulting design (as shown by Lee & Pennington, 1994, for software design using an object-oriented or a procedural paradigm).

*Dependencies Between Function and Form.* An initial distinction opposes domains where function and form can be aligned, to domains where individual forms are devised to do many functions simultaneously. In the first type of domain, to each particular form corresponds a particular function. In software design (e.g., to a functional decomposition corresponds directly form decomposition). In the second type of domain (e.g., mechanical design) each design decision can affect each subsequent decision, because a goal may be achieved by modifying a previously specified form rather than by introducing a new one (Ullman et al., 1988). One may suppose that these differences imply differences in decomposition activities.

*"Designing in Space Versus Time" (Thomas & Carroll, 1979/1984).* Research has shown that designers deal differently with temporal and spatial constraints (Chalmé, Visser, & Denis, 2004; detailed in Visser, 2004), but the relative difficulty and especially the underlying factors of these two types of design have not been settled unambiguously.

*Distance Between Design Concept and Final Product.* Löwgren (1995,
p. 94) opposes "external" software design ("design of the external behavior and appearance of the product, the services it offers to users and its place in the organization") to other types of design, e.g., architectural and engineering design. In external software design "it is technically possible to evolve a software prototype into a final product"—something difficult (or even impossible) in these other fields. Therefore, in domains such as external software design, "the 'distance' between the design concept and the final product is shorter than in, say, architecture" (p. 93). This does not imply, however, that in those domains, design and implementation are not separated. It might, however, clarify our observation that software designers find it particularly difficult to separate design from coding (Visser, 1987).

*Delay of Implementation.* Something that has been considered to make the solving of social-science problems quite difficult, is the "delay from the time a solution is proposed and accepted to when it is fully implemented" (Voss et al., 1983). "Naturally, a good solution anticipates changes in conditions, but anticipation can be quite difficult" (Voss et al.). This remark that was formulated with respect to social-science problems is particularly applicable in social domains, but may also hold for other design areas. The underlying factor may influence evaluation of solution proposals.

*Possible Forms of Evaluation.* Domains differ in the means that may be used to evaluate design proposals (Malhotra, Thomas, Carroll, & Miller, 1980, pp. 129–130). In engineering, "objective" measures of future artefacts' performance can be used. One can calculate whether a particular design meets functional requirements (e.g., accommodation and maximum load). Using such measures, different proposals can be ranked somewhat objectively. The results of qualitative evaluation on subjective criteria such as aesthetics, are much more difficult to translate into a "score."

*Artefacts' Behaviour Over Time.* "Interactive systems are designed to have a certain behavior over time, whereas houses typically are not" (Löwgren, 1995, p. 94). Even if this assertion is questionable, their behaviour over time is a dimension on which artefacts differ. Houses may not display "behaviour" over time, but they change. "Good" designers anticipate the transformations that their design may undergo through, for example, its use. Nevertheless, systems such as organisations are subject to stronger evolutions than buildings.

*Artefacts' Transformative Nature.* Artefacts' behaviour over time is to be distinguished from what Carroll et al. (1997, p. 63) call the "transformative" nature of certain systems. The systems Carroll and al. develop (enabling novel educational activities) "fundamentally [alter] possibilities for human behavior and experience."

*Other Dimensions.* Other possible dimensions are: *design of structures vs. design of processes*, *role of the user in design*, *maturity of a domain*, *impact of an artefact on human activity and the possibility to anticipate it*. "Software design versus other 'forms' of design" is a label in search of underlying dimensions. One frequently encounters in the cognitive design research literature allusions to, or implicit testimonies of the specific character of software design compared to other types of design (Atwood, McCain, & Williams, 2002). The responsible dimensions remain, however, unexplored.

Our list of candidate dimensions that might differentiate forms of design is a start for their discussion and analysis. A further step would be to elucidate if indeed, and if so how these and other differences influence design activity and its result, the artefact.

Linking Representational Structures and Activities

This section examines functional linkages between representational structures and activities through the scheme "Input representation–Activity–Out-put representation". Different dependencies exist: the input or output representation may constrain an activity (I_Repr or O_Repr _ Act), or be constrained by the activity (Act _ I_Repr or O_Repr), or two elements can mutually constrain each other (_). The constraint may concern different aspects of an element. For representations, it may, for example, be their external or internal, notational or non-notational nature (in Goel's 1995 terms).

Our first observation is a case of Act _ I_Repr, Act constraining the notational nature of I_Repr. We suppose that activities concerned with gaining new perspectives on representations (association, [re]interpretation, inductive inference) will more often be based on non-notational representations, such as sketches and other fluid forms of representation, than on notational representations, such as drafts with dimensions and tolerances.

The second case is one of Act _ O_Repr, where Act constrains the inter-nal–external character of O_Repr. Interpretational activities will specifically result in internal representations —even if designers may use external representations to proceed to these representational activities.

The provisionality that non-notational forms of drawings may convey through roughness, and face-to-face, oral communication through specific phrasing and intonation, points to an Act _ I_Repr relation between these non-notational representational systems and particular interactional activities in cooperative design.

The last two examples concern I_Repr _ Act. First, in domains of design concerned with physical artefacts, a common association exists between sim-ulation-based evaluation and external representations (from two-dimensional drawings to three-dimensional mock-ups). Second, calculation and comparable activities will use notational, more formalised representational systems and will, in design, be associated with particular notational types of representations such as drafts with dimensions and tolerances. These algorithmic activities may be used for the quantitative form of evaluation.

The next step toward a process model of design will require identifying more such functional linkages, and, especially, the organisation of the different activities and corresponding representations, into a structure, such as the blackboard framework that we adopted in previous studies for the designers' organisation of their activity (Visser, 1994).

## 4. CONCLUSION

In this conclusion, we will first come back to issues that have been central to this article: SIP, SIT and design, and design as problem solving. We then onclude our article with a brief review of the proposal we have been intro ducing in this article.

## 4.1. SIP and SIT, Individual and Collective Design

As noticed several times in this article, our critique on the SIP paradigm concerns its oversystematic view of design. Where Simon seems to focus on the common points between design and other problem-solving tasks (his "nothing special" view), Schön and other SIT researchers seem to be more interested in design's specificities.

Greeno (1997), once collaborating with Simon (Greeno & Simon, 1988), came to adopt a SIT perspective. He judges that the SIP perspective

Takes the theory of individual cognition as its basis and builds toward a broader theory by incrementally developing analyses of additional components that are considered as contexts. The situative perspective takes the theory of social and ecological interaction as its basis and builds toward a more comprehensive theory by developing increasingly detailed analyses of information structures in the contents of people's interactions. (p. 5)

This viewpoint is compatible with our view that the two positions are not contradictory, but complementary—not only because SIP would focus on individual design, and SIT on interactions in design.

Nevertheless, the switch of focus in design studies, from individual to collaborative design, has entailed an evolution of the theoretical frameworks. The purely cognitive framework based on the SIP approach is not sufficient for modelling individual design, but to address the collective, interactional nature of work, it is certainly inadequate. Because of its interest for situational resources, the SIT approach has *in principle* the potential to propose a more appropriate view on design. We have seen in this article how recent studies adopting the reflective-practitioner viewpoint are developing increasingly detailed analyses of central SIT notions—and not only in collective design settings.

## 4.2. Design Is Not Problem Solving: Design Involves Problem Solving

Many authors state that design is not "problem solving." Others consider that design is *not only* problem "solving" but also problem "setting," "structuring," and "framing" to name just a few other activities advanced. Certain criticisms addressed by SIT proponents at the SIP approach are based, in our view, on terminological confusions, especially concerning "problem" and "problem solving." For a naïve reader, these terms may indeed seem inappropriate when applied to design. As we noticed however, even if the technical cognitive-psychology acceptation of "problem solving" covers a wide range of problem-centred activities, many of them, especially the construction and use of representations, do not receive much attention in SIP-oriented studies. Nevertheless, these two activities are at the core of design. That is why, in our view, even if design does indeed *involve* problem solving in a broad sense, characterising it as "problem solving" does not capture its essence! Design is problem solving in the sense that its requirements generally will not evoke a memorized procedure, but this does not tell us much about the activities used to "solve" the corresponding "design problem"!

## 4.3. Design as Construction of Representations

Most features presented in this article as characteristic of design contribute to characterise the underlying activity as being multifaceted: its ill-definedness, complexity, ambiguity, the incomplete and especially the conflicting nature of its constraints—and the importance of representations, diverse with respect to their abstraction and precision, their internal or external, notational or non-notational character.

In an activity that functions by way of representations, knowledge plays a central role. Recognition is important, too—and Simon was completely right in highlighting its role. Yet, we consider that Simon overestimated its importance relative to the controlled use of knowledge. To recognise a potentially relevant element of knowledge, there must be memory associations between target (features of the situation) and source (knowledge elements). Otherwise, controlled search or exploration has to be used, leading—if one is lucky—to discoveries of unintended, useful consequences.

As regards a professional activity as design, the knowledge on which both recognition and analogical reasoning and other knowledge-intensive activities are based, is grounded mainly in professional experience (but not only; see Visser, 1995b). Such knowledge also shapes the ill-definedness of a problem. For a designer with extensive and long experience in several application domains, a task that constitutes an ill-defined problem for an inexperienced colleague, may constitute a task with routine aspects. Knowledge-intensive tasks have an evolving nature with respect to their "problem" character, their "routine" character, their "ill-definedness"—and possibly even their "design" character.

Adopting an entirely "problem-solver-oriented" approach, Thomas and Carroll (1979/1984, p. 222) proposed viewing design, not as a particular "type" of problem, but as a "way of looking at" a problem—we might say a particular representation of a problem. Theorem proving, for example, can be viewed as designing if, say, the requirement to stay inside certain formal rules is relaxed and creativity is allowed; in this case, although the initial goal was to prove a theorem, the goal in fact becomes "to find out something interesting." In the same way, "designing" a house by applying a set of standard rules to the stated requirements is no longer "design" in a cognitive-activity oriented sense. Therefore, whether a task is "design" depends entirely on the person who faces the task. "Much of what [people] call technological progress may be viewed as a process of rendering ill-structured design problems as more well-structured procedures for accomplishing the same ends—without requiring design" (Thomas & Carroll, 1979/1984, p. 222). Indeed, if a design task is no longer open-ended, ill-defined, ambiguous, if its constraints are the object of agreement, a "design problem" can become a "transformation prob-lem"—or even no longer constitute a "problem" at all!

Our current proposal is consequential in that it opens new directions, both for research and for support. Qualifying design as problem solving does not inform us of the activities and structures implemented. The qualification of design as construction of representations guides the further exploration of forms that this construction may take, through still other activities and representational structures, depending on still other dimensions, than we have proposed in this article.

A new vision of a domain provides new possibilities for its study, and as a consequence opens new views of needs for assistance and the potential support modalities. If domain-specific but varied knowledge, and representational activities and structures are indeed so essential in design, if designers' way of looking at their projects is indeed so critical to their success, then a valuable approach is the development of systems—not only technological, but also methodological and organisational—that may support these activities and structures, and encourage the development of such knowledge through designers' involvement in many different types of projects.

## NOTES

*Acknowledgments.* The author wishes to thank Françoise Détienne, Pierre Falzon, and four anonymous reviewers for their precious comments on earlier versions of this article. They obliged me to go further in my reflections on design and construction of representations, on design *as* construction of representations.

*Author's Address.* Willemien Visser, INRIA—National Institute for Research in Computer Science and Control, EIFFEL—Cognition & Cooperation in Design, Bât. 23 -Rocquencourt B.P. 105, 78153 Le Chesnay Cedex, France. Email: Willemien.Visser@inria.fr.